\documentclass[prl,preprint,onecolumn,showpacs,letterpaper]{revtex4}

\usepackage{graphicx}

\begin{document}

\title{How Hertzian solitary waves interact with boundaries in a 1-D granular medium}

\author{St\'ephane Job}
\altaffiliation{Permanent adress: {\sc Supmeca} - 3, rue Fernand Hainaut 93407 Saint-Ouen Cedex - France}
\email{stephane.job@supmeca.fr, http://www.supmeca.fr/perso/jobs.}
\author{Francisco Melo}
\affiliation{
Departamento de F\'{\i}sica, Universidad de Santiago de Chile,\\
and Center for Advanced Interdisciplinary Research in Materials (CIMAT),\\
Av. Ecuador 3493, Casilla 307, Correo 2, Santiago de Chile.}

\author{Adam Sokolow}
\author{Surajit Sen}
\affiliation{
Department of Physics,
State University of New York at Buffalo,
Buffalo, New York 14260, USA.
}

\received{11 January 2005}
\published{3 May 2005}

\begin{abstract}
We perform measurements, numerical simulations, and quantitative
comparisons with available
 theory on solitary wave propagation in a linear chain of beads without static preconstrain. By designing
 a nonintrusive force sensor to measure the impulse as it propagates along the chain, we study the solitary
  wave reflection at a wall. We show that the main features of solitary wave reflection depend on wall
   mechanical properties. Since previous studies on solitary waves have been performed at walls without
    these considerations, our experiment provides a more reliable tool to characterize solitary wave propagation.
     We find, for the first time, precise quantitative agreements.\\\\
     DOI: 10.1103/PhysRevLett.94.178002
\end{abstract}


\maketitle

Solitons are widely studied in physics because of their ubiquity in
systems exhibiting nonlinear propagation~\cite{Drazin2002}. In a
granular chain, theoretical and experimental evidence of
solitons  was first reported
 by Nesterenko~\cite{Nesterenko1984,Lazaridi1985,Gavrilyuk1994,Nesterenko1994,Nesterenko1995,Nesterenko2001}.
 Since Nesterenko's  pioneering work, most of the experimental effort in the field has generally
 focused on the scaling laws for amplitude and speed of the
 solitons,~\cite{Coste1997,Coste1999}.  It was recently reported~\cite{Manciu2001a} that
  identical and opposite propagating solitons do not preserve themselves
  upon collision and hence these are solitary waves rather than
  solitons.  Several detailed numerical studies have been devoted to
  understand the  interactions of solitary waves with a perfectly reflecting
  wall~\cite{Manciu2001b,Manciu2001a,Manciu2002,Xu2001,Hascoet1999}
   and show that tiny secondary solitary waves are generated as a solitary wave
   is reflected off a wall~\cite{Manciu2001a,Manciu2002}.  However,
due to experimental difficulties, no close comparison between
experiments and simulations has so far been established. Here
inspired by Nesterenko's
experiments~\cite{Nesterenko1994,Nesterenko1995,Nesterenko2001}, we
developed an adapted impulse sensor to nonintrusively investigate
solitary wave propagation in a linear chain of identical elastic
beads. We explored the problem of solitary wave reflection by
changing the elastic properties of the wall and showed that the
solitary wave detected at the wall differs from the actual solitary
wave propagating through the chain. Our measurements
significantly improve upon previous experimental studies
~\cite{Lazaridi1985,Coste1997} and allows excellent agreement with
our numerical simulations and Nesterenko's analytical theory
~\cite{Nesterenko2001}.

The physical behavior of solitary waves in bead chains can be
described as follows. Under elastic deformation, the energy stored
at the contact between two elastic bodies submitted to an axial
compression corresponds to the Hertz potential~\cite{Landau1967}
$U_{H}=(2/5)\kappa\delta^{5/2}$, where $\delta$ is the overlap
deformation between bodies,
$\kappa^{-1}=(\theta+\theta')(R^{-1}+R'^{-1})^{1/2}$,
$\theta=3(1-\nu^2)/(4Y)$, and $R$ and $R'$ are radii of curvature at
the contact. $Y$ and $\nu$ are Young's Modulus and Poisson's ratio,
respectively. Since the force felt at the interface is the
derivative of the potential with respect to $\delta$,
($F_{H}=\partial_\delta U_H=\kappa\delta^{3/2}$), the dynamics of
the chain of beads is described by the following system of $N$
coupled nonlinear equations,
\begin{equation}\label{Eq:DiscreteEquation}
m\partial_{tt}^2u_n=\kappa\left[(u_{n-1}-u_{n})^{3/2}_{+}-(u_{n}-u_{n+1})^{3/2}_{+}\right],
\end{equation}
where $m$ is the mass, $u_n$ is the position of the center of mass of bead $n$,
$u_n=2nR$ at rest, and the label $+$ on the brackets indicates that
the Hertz force is zero when the beads lose contact. Under the
long-wavelength approximation $\lambda\gg R$ (where $\lambda$ is the
characteristic wavelength of the perturbation), the continuum limit
of Eq.~\ref{Eq:DiscreteEquation} can be obtained by replacing the
discrete
 function $u_{n\pm1}(t)$ by the Taylor expansion of the continuous function
  $u(x\pm2R,t)$. Keeping terms of up to the fourth order spatial
  derivatives, Eq.~\ref{Eq:DiscreteEquation} leads to the equation
  for the strain $\psi=-\partial_x u>0$,
\begin{equation}\label{Eq:ContinuousEquation}
\partial_{tt}^2\psi\simeq c^2\partial^2_{xx}[\psi^{3/2}+(2/5)R^2\psi^{1/4}\partial^2_{xx}(\psi^{5/4})],
\end{equation}
where $c=(2R)^{5/4}(\kappa/m)^{1/2}$ ~\cite{Nesterenko2001}. Looking
for progressive waves with speed $v$, in the form $\psi(\xi=x-vt)$,
Eq.~\ref{Eq:ContinuousEquation} admits an exact periodic solution in
the form $\psi=(5/4)^2\times(v/c)^4\cos^4[\xi/(R\sqrt{10})]$
\cite{Nesterenko1984,Lazaridi1985,Gavrilyuk1994,Nesterenko1994,Nesterenko1995}.
Although this solution only satisfies the truncated
Eq.~\ref{Eq:DiscreteEquation}, there is quantitative analysis on
how well one hump ($-\pi/2<\xi/(R\sqrt{10})<\pi/2$)
 of this periodic function represents a soliton solution ~\cite{Nesterenko1984,Chatterjee1999}.
  Approximating the spatial derivative, the strain in the chain reads $\psi\simeq\delta/(2R)$,
  and the force felt at beads contact,
  $F\simeq\kappa(2R\psi)^{3/2}$, and $v$ become,
\begin{equation}\label{Eq:SolutionForce}
F\simeq F_m\cos^6{\left[\frac{x-vt}{R\sqrt{10}}\right]};
v\simeq\left(\frac{6}{5\pi\rho}\right)^{\frac{1}{2}}\left(\frac{F_m}{\theta^{2}
R^{2}}\right)^{\frac{1}{6}}.
\end{equation}

\begin{figure}[t]
\includegraphics[width=16cm]{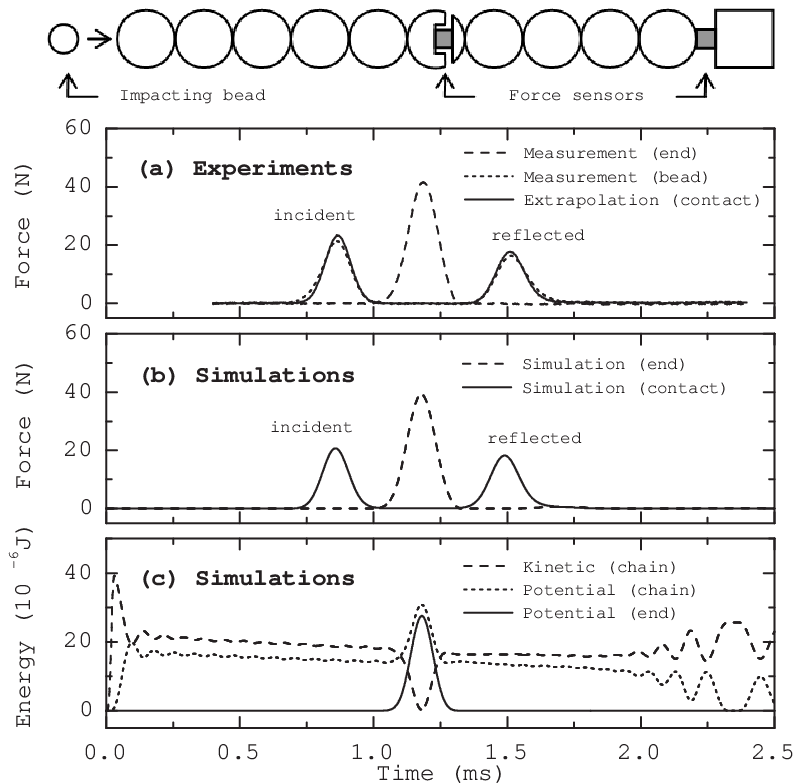}
\caption{\label{fig1} (top)
Schematic view
 of experimental setup. (a) Experiments: middle peak indicates the force
 signal at the end of the chain, whereas lateral peaks are the incident and reflected solitary
  wave. The solid line represents the force at a single contact extrapolated from
  Eq.~\ref{eq:SensorForce}.
  (b) and (c) The numerical simulations of the contact forces and energy, respectively, for $\eta_{n}=2\mu$s.}
\end{figure}

\begin{figure}[t]
\includegraphics[width=16cm]{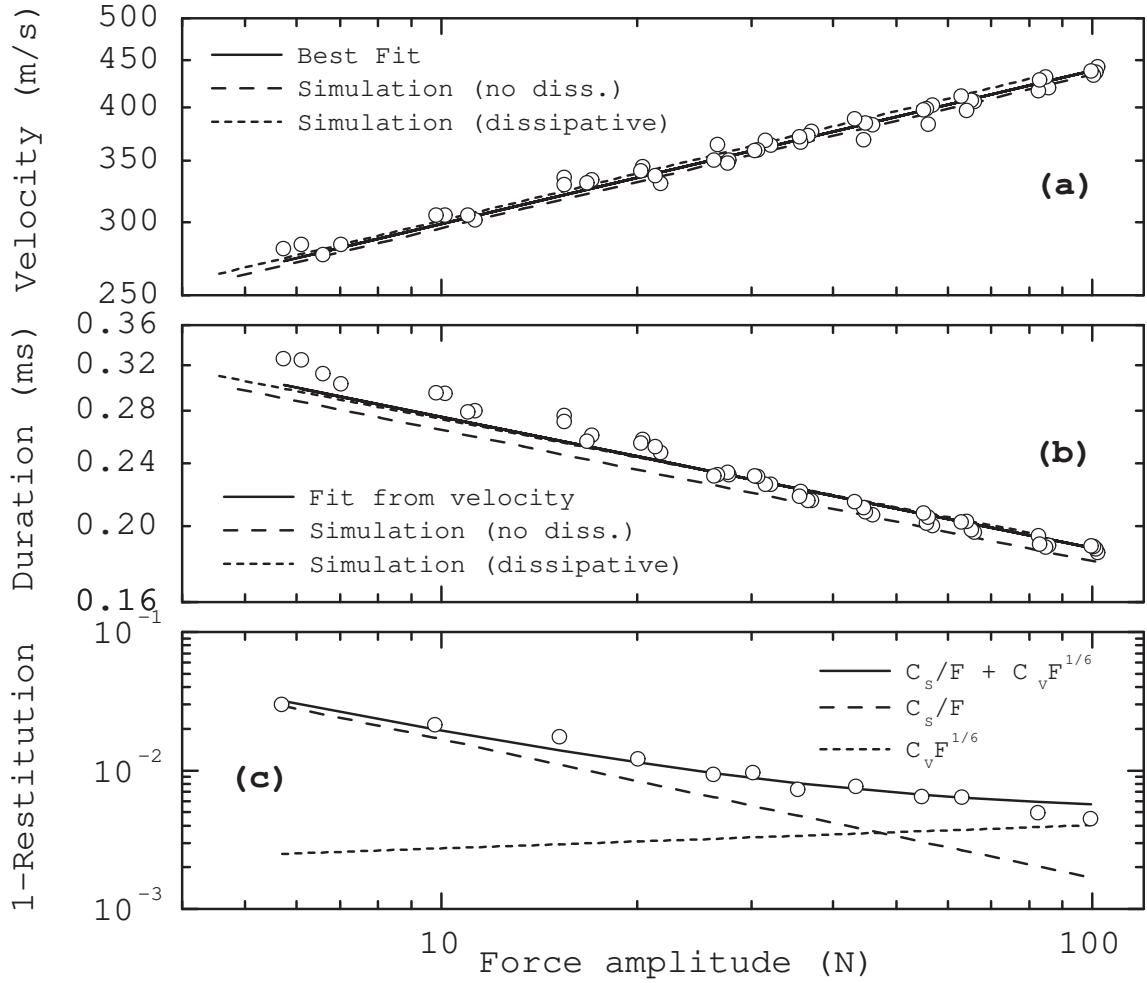}
\caption{\label{fig2} (a) Velocity $v$ and (b)
 duration $2\tau$ of the solitary wave, measured inside the chain, vs
  force amplitude. Theoretical predictions from Eq.~\ref{Eq:SolutionForce},
  and numerical simulations are contrasted to experimental data.
   (c) Restitution coefficient vs force amplitude.}
\end{figure}

In our experiment, we consider the chain of $21$ identical beads of
mass $m$, located on a Plexiglas linear track as shown on top of
 Fig.~\ref{fig1}. A piezoelectric dynamic impulse sensor
 ({\em PCB 208A11} with sensitivity $112.40$~mV/N) located at the end of the chain provides
  the force at the rigid
  end. This sensor has a flat cap made of the same material as the beads. Beads are
   {\em Tsubaki} high carbon chrome hardened steel roll bearing (norm {\em JIS SUJ2} equivalent
   to {\em AISI 52100}). The radius of the beads is $R=13$~mm (tolerance is $\pm125$~$\mu$m on diameter),
    and the density is $\rho=7780$~kg/m$^3$. The Young's modulus is $Y=203 \pm4$~GPa~\cite{Tsubaki},
    and the Poisson ratio is assumed to be $\nu=0.3$; our beads have thus a $\kappa=12$~N/$\mu$m$^{3/2}$.
     Moreover, the deformation keeps elastic and below yield stress ($\sigma_Y=2$~GPa~\cite{Tsubaki}).
     Assuming that the contact surface is a disk of area $A=\pi({\theta}RF)^{2/3}$~\cite{Landau1967},
    the corresponding maximum compression force is roughly $F_Y\simeq470$~N, which corresponds
     to an overlap $\delta_Y\simeq11$~$\mu$m.
Forces inside the chain are monitored by a flat dynamic impulse
sensor ({\em PCB 200B02} with sensitivity $11.24$~mV/N) that is
inserted inside one of the beads, cut in two parts. The total mass
of the bead sensor system has been compensated to match the mass of
an original bead. This
 system allows achieving non intrusive force measurement by preserving both
 contact and inertial properties of the bead-sensor system. The stiffness of the
  sensor $k_s=1.9$~kN/$\mu$m being greater than the stiffness of the Hertzian contact
  ($k_s\gg k_{H}\propto \kappa\delta^{1/2}$), means the coupling between the chain and the sensor
   is consequently negligible. To relate the force $F_s$ registered by the sensor with
   the actual force at the beads contact, we write
    the Newton's law for both masses, respectively, located
   in front ($+$) and in the back ($-$) of the sensor.  Thus,
   $F_\pm=F_s\pm m_\pm\partial_{tt}^2x_\pm$, with $F_s=k_s(x_+-x_-)$.
   This set of equations can be summarized as
\begin{equation}\label{eq:SensorForce}
\partial_{tt}^2F_s+\omega_0^2F_s = \omega_0^2\left[(1-\beta)F_+ +
{\beta}F_-\right],
\end{equation}
where we have introduced the resonant angular frequency of the system $\omega_0=[k_s(m_+^{-1}+m_-^{-1})]^{1/2}$,
 and the mass ratio $\beta=m_+/(m_++m_-)$.
 Experimentally $\beta=0.11$ and
 the resonant frequency,
  $f_0=\omega_0/(2\pi)\simeq85$~kHz, indicates that safe measurements can be obtained for signal
  whose period is greater than $\tau_0=1/f_0\simeq12$~$\mu$s. However, a relation between
  $F_\pm(t)$ is needed to invert Eq.~\ref{eq:SensorForce} and then determine the force
   $F_+(t)$ or $F_-(t)$ from the force $F_s(t)$. Assuming that the pulse
   travels at a velocity $v$, this relation reads $F_-(t)=F_+(t+t_0)$, where $t_0=(x_+-x_-)/v$.
   An estimate of the velocity $v$
   is obtained from the time of flight of the pulse and the deconvolution
    of Eq.~\ref{eq:SensorForce} by means of Fast Fourier Transform,
   then provides the actual force $F_+(t)$ felt exactly at the interface between two beads.
   Notice that the improvement introduced here represents a correction of the order of $\beta$,
    i.e. about $10\%$.
Signals from sensors are amplified by a conditioner ({\em PCB
482A16}), recorded by a two channels
 numeric oscilloscope ({\em Tektronix TDS340}), and transferred to a computer.
 The acquisition is triggered by the contact between
  the small impacting bead and the chain; both being in contact with soft wires they
  cause the discharge of a capacitor in a resistor ($1/RC\simeq1$~$\mu$s). This circuit
  allows high repeatability, e.g. for time of flight
  measurements. In Fig.~\ref{fig1}a,
  a solitary wave propagates along the chain of beads. The central peak corresponds
   to the impulse detected at the end, whereas the two peaks on the sides are the
   incident
   and reflected waves measured inside the chain. Notice that the central peak
   is much higher and broader than the actual solitary wave propagating along the
   chain, thus no quantitative information can be extracted from it without a
   detailed description of the interaction between the solitary wave and the wall. In order
    to characterize solitary waves, we look both for velocity and duration of incident
     pulses recorded  at one contact far from the wall. According
      to Eq.~\ref{Eq:SolutionForce}, we map experiments to $F(t)=F_m\cos^6[(t-t_0)/\tau]$,
       to obtain the amplitude $F_m$, the duration $2\tau$, and the time of flight $t_0$ of
        a pulse. To provide more accurate data for the velocity, we perform
           flight time measurements for different positions of the active bead. In addition, for every
             experimental configuration we record three sets of data and check repeatability,
             and the whole experiment is repeated three times. According to Eq.~\ref{Eq:SolutionForce}, we first look for the best fit
    in a least squares sense for the experimental velocity of the pulse, in the form
    $v=CF_m^{1/6}$, and we find an experimental value $C_{e}=203.6\pm8.9$ in standard units.
    This value agrees with the theoretical prediction $C_{t}=198.9$, derived from Eq.~\ref{Eq:SolutionForce},
     within an error less than $3\%$. The fit is plotted in straight line in
     Fig.~\ref{fig2}a. For sake of comparison, we also plot
      (the straight line on Fig.~\ref{fig2}b) the duration
      $2\tau=2R\sqrt{10}/v$, also obtained from Eq.~\ref{Eq:SolutionForce}. The velocity is thus in
      a satisfactory agreement with the theoretical prediction, which also appears at first glance
      to predict in a good manner the duration of the pulse. However, energy dissipation is expected
      to produce a broader solitary wave. Dissipation is characterized by the restitution coefficient
      (see Fig.~\ref{fig2}c) defined as
      $\epsilon=(U_{n+1}/U_{n})^{1/2}=(F_{n+1}/F_{n})^{5/6}$ ($U_{n}$ is the Hertz potential, i.e.,
      the work done by the Hertz force $F_{n}$ at the contact $n$). Here we consider two mechanisms responsible for
      the dissipation; internal viscoelasticity and solid friction of beads
      submitted to their weight $mg$ ($g$ is the gravity), on the track.
A third mechanism, the solid friction between
      beads due to thwarted rotations~\cite{Duran1997}, may also be taken into account. However, the
      contribution of a friction force of the form $F_{s}^{*}=\mu^{*}\kappa\delta^{3/2}$ into
      Eq.~\ref{Eq:DiscreteEquation} reduces simply to considering an equivalent nonlinear stiffness
      $\kappa^{*}=(1+\mu^{*})\kappa$.
Viscoelastic dissipation is included by using the simplest
       approximation~\cite{Kuwabara1987,Brilliantov1996} for which the dissipative force at the contact
        of two beads reads, $F_{v}=\eta\kappa\partial_t(\delta^{3/2})$, where $\eta$ includes unknown
         coefficients due to internal friction of the material~\cite{Landau1967,Brilliantov1996}.
         Solid friction is taken into account by considering a frictional
         force $F_{s}=\mu mg$~\cite{Duran1997}. The potential energy difference $(U_{n}-U_{n+1})$
          being equal to the work done by both previous dissipative forces allows us to estimate the
          restitution coefficient to be force dependent, 
           $\epsilon=(U_{n+1}/U_{n})^{1/2}\simeq1-C_{v}F^{1/6}-C_{s}/F$.
          Simple calculations provide the relation of $\eta$ and $\mu$ with the new
           constants $C_{v}$ and $C_{s}$
           as,  $\eta \approx{ 2\sqrt{10}RC_{v}/5C}$ and $\mu \approx{4C_{s}/5mg}$ respectively.
          Experimentally, we determine that $C_{v}=1.9\times10^{-2}$ and $C_{s}=1.7\times10^{-1}$
          in standard units, see
          Fig.~\ref{fig2}c.  Then,
          $\eta_{e}\approx{1.8}$ $\mu$s and $\mu_{e}\approx{0.19}$.

         Numerical simulations based on a Velocity-Verlet algorithm
          allow to explore the main features of solitary waves by solving
           Eq.~\ref{Eq:DiscreteEquation} directly. We first run numerical calculation without
          dissipation, plotted in dashed lines on
           Fig.~\ref{fig2}a and \ref{fig2}b.
           Looking for least square fit for the velocity, as previously done, we find $C_{n}=201.5\pm0.1$.
            Compared to the theoretical value $C_t$, simulations improve the agreement with experiments
            (relative error on velocity is about $1\%$), but a noteworthy disagreement is now observed
            for the duration of the pulse (see Fig.~\ref{fig2}b), which
            is about $10\%$ lower than experimental values. This lag is consistent with the presence of a
            weak dissipation. At this stage, we only consider the effect of viscoelastic dissipation in
            numerical simulations. We thus adjust the coefficients, and for $\eta_{n}=2$~$\mu$s and $\mu=0$,
            a good agreement can be obtained both for the velocity and the duration, in the range of
            amplitude where viscoelastic dissipation dominates over solid friction
            ($F_m>50$~N).  Notice that $\eta_{n}$ differs from the experimental value $\eta_{e}$ only by $20\%$.
            Since solid friction has not yet been included in
            simulations, the experimental pulse is still broader than
              in simulations at low force amplitude ($F_m<20$~N)
              where this mechanism dominates.

We now check how simulations reproduce the features of the
reflection process. Figure~\ref{fig1}b
shows the corresponding numerical simulations for the incident and the
reflected solitary wave as well
 as the force registered at the wall. Although simulations include only viscous dissipation, $\eta_{n}=2$~$\mu$s,
 the agreement between
 Fig.~\ref{fig1}a and
 Fig.~\ref{fig1}b
is very good. Notice that momentum is conserved,
 i.e., the area of the central peak in Fig.~\ref{fig1}a
  is  equal to the area of the incident plus the reflected solitary wave.
    Figure~\ref{fig1}c presents the corresponding calculations of
    the time evolution of the
     potential and kinetic energy when a solitary wave interacts with the wall sensor. The solitary
     wave is initiated
     at $t=0$ by a purely kinetic impact. At $t=1$~ms the pulse reaches the rigid
     sensor and the energy is stored into potential. The pulse is then reflected
      and propagates backward to the free end until leading to ejection of beads after $t=2$~ms.

\begin{figure}[t]
\includegraphics[width=16cm]{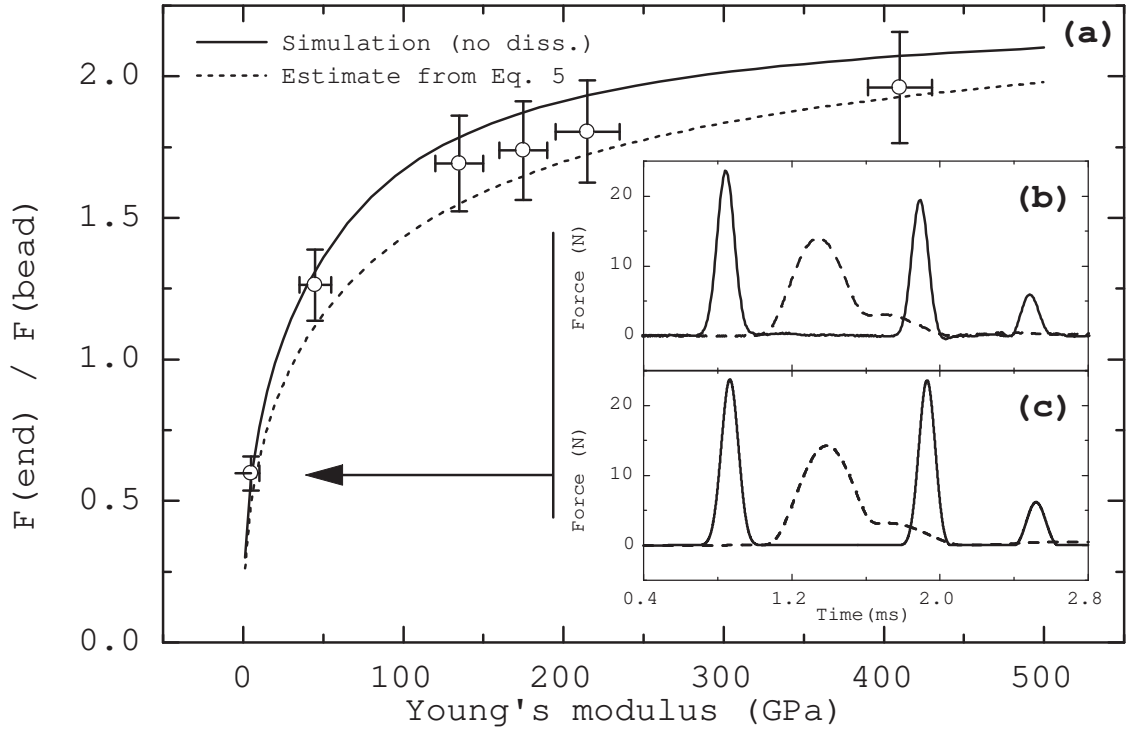}
\caption{\label{fig3} (a) Ratio of force
 amplitudes at the end of the chain and of the incident impulse vs
 Young's modulus of the sample placed on the rigid sensor. Inset (b) force
  measurements when a solitary wave collides on the softer sample ($Y=5$~GPa), and (c)
   corresponding simulation reproducing all the experimental features for $\eta_{n}=0$. Dashed lines indicate forces at the end
   of the chain.  The last peaks
    on the right represent the secondary solitary waves.}
\end{figure}

We further investigated the solitary wave reflection by varying the mechanical
 properties of the flat part of sensor in contact with the last bead. This is
 done by locating polished disks of $1$~mm thickness and $5$~mm diameter of
 different known materials on the active part of the sensor. These samples
 are made of plexiglass, {\em Mg}, {\em Cu}, {\em Si}, {\em Fe}, and {\em W}.
 For materials softer than the beads, unexpected features arise.
 For instance, in Fig.~\ref{fig3}b,
 the experimental force on the wall exhibits a well defined secondary peak.
  The break of symmetry implied by the change of elastic properties leads to the
   generation of a so-called secondary solitary wave in the reflected impulse
   predicted recently via simulations in~\cite{Manciu2001b,Manciu2002}. Dissipationless numerical
    simulations in Fig.~\ref{fig3}c reproduce
    well the experimental finding of Fig.~\ref{fig3}b without
    adjustable parameter.  Better agreement can be achieved but it requires
    the knowledge of the mechanism dominating dissipation of the samples.
Fig~\ref{fig3}a
    is the ratio of the maximum force measured at the wall and the respective maximum
     force of the incident solitary wave.  Despite the peculiar form of the force, the
      ratio of maximum forces follows a well defined law that is
       characteristic of the kinetic to potential energy conversion at the
       wall. This interesting feature should prove valuable to determine the Young
        modulus of materials of unknown nature, when the sample size is a practical
         limitation.

To understand the underlying physics of solitary wave reflection, we
focus on the kinetic-potential
 energy conversion when a solitary wave interacts with a rigid wall. As shown
 on Fig.~\ref{fig1}c and~\cite{Nesterenko1984,Nesterenko2001}, when solitary wave propagates
  freely in the chain, the kinetic energy $K_{chain}$ is about 56\% and the potential
   energy $U_{chain}$ is about 44\% of the total energy (for a rough estimation we
   assume $K_{chain}\simeq U_{chain}$). However, when a solitary wave reaches the end
    of the chain, the potential energy stored at the sensor-bead contact equals the
    total energy carried by the solitary wave. The kinetic energy is thus transformed
     into potential at the contact. Then, $U_{end}^{max}\simeq2U_{chain}$. On the other
      hand, the solitary wave extends on a few beads, and the potential
      energy stored in the chain is roughly supported by the most compressed contact
       ($U_{chain}\simeq U_{bead}^{max}$). It finally becomes,
\begin{equation}\label{eq:ForceRatioEstimation} \frac{U_{end}^{max}}
{U_{bead}^{max}}\simeq2\rightarrow\frac{F_{end}^{max}}{F_{bead}^{max}}\simeq
2^{6/5}\left(1+\frac{Y_{bead}}{Y_{end}}\right)^{-2/5}
\end{equation}
which is a function of Young modulus  of beads and the sensor plane.
In Fig.~\ref{fig3}a, we compare
experiments, numerical simulation, and the above estimate. Within the
error bars, a satisfactory agreement is obtained.

In conclusion, we have developed a non intrusive reliable method to
investigate solitary wave propagation and solitary wave reflection
at walls.
 Our measurements in conjunction with our numerical simulations provide a powerful
  tool to accurately investigate a variety of related problems such as
  the main features of solitary waves reaching impedance mismatch, the generation of
   the recently predicted secondary solitary waves at the boundaries, and the solitary wave
    interactions, among others.

This work received the support of {\sc Conicyt} under program {\sc
Fondap} No. 11980002. The Consortium of the Americas is acknowledged
for supporting the visit of A.S. to Chile. S.S. acknowledges partial
support of NSF-CMS-0070055.

\bibliography{Soliton_Job}


\clearpage

\clearpage

\clearpage

\clearpage

\end{document}